\documentclass[11pt]{article}
\usepackage{amsfonts}
\usepackage{epsfig}

\def\underset#1#2{\mathrel{\mathop{#2}\limits_{#1}}}

\newfont{\blackb}{msbm10 scaled\magstep1}

\newfont{\calig}{cmsy10 scaled\magstep1}

\def\text#1{\hbox{#1}}

\newtheorem{theorem}{Theorem}[section]

\newtheorem{remark}{Remark}[section]

\def\be{\begin{equation}}
\def\ee{\end{equation}}
\def\ben{\begin{displaymath}}
\def\een{\end{displaymath}}
\def\baa{\begin{eqnarray}}
\def\eaa{\end{eqnarray}}

\def\ba{\begin{array}}
\def\ea{\end{array}}

\def\g{\gamma}
\def\xb{{\bar{\xi}}}
\def\E{{\cal E}}
\def\Eb{\bar{{\cal E}}}
\def\3{\ss}

\def\ka{\kappa}
\def\l{\lambda}

\def\s{\sigma}
\def\t{\tau}

\def\Th{\Theta}

\def\O{\Omega}
\def\x{\xi}
\def\phi{\varphi}
\def\Qt{\tilde{Q}}
\def\Rt{\tilde{R}}
\def\eb{{\bf e}}

\def\Lh{\hat{\L}}
\def\kat{{\tilde{\ka}}}

\def\B{{\bf B}}
\def\C{\mathbb{C}}
\def\Z{\mathbb{Z}}
\def\R{\mathbb{R}}

\def\t0{\Theta_0}
\def\z{{\bf z}}
\def\m{{\bf m}}
\def\la{\label}

\def\c{\cite}
\def\f{\frac}
\def\L{{\cal L}}
\def\p{\partial}
\def\pb{{\bf p}}
\def\qb{{\bf q}}

\def\rb{{\bf r}}
\def\sb{{\bf s}}

\def\tr{{\rm tr}}
\def\0{S}
\def\1{T}

\def\log{\ln}
\def\ol{\overline}
\begin{document}

\begin{center}{\LARGE
Solutions of Schlesinger system and
Ernst equation in terms of theta-functions}\\
\vskip1.0cm
{\large D.~A.~Korotkin}\footnote{E-mail: korotkin@aei-potsdam.mpg.de}\\
\vskip0.5cm
Max-Planck-Institut f\"ur Gravitationsphysik, \\
Schlaatzweg 1, D-14473 Potsdam, Germany\\
\vskip0.5cm
and\\
\vskip0.5cm
{\large V.~B.~Matveev} \footnote{E-mail: matveev@u-bourgogne.fr}\\
%\addtocounter{footnote}{-1}
\vskip0.5cm
Physique Math\'{e}matique, UFR Sciences et Techniques,\\ 
Universit\'{e} de Bourgogne,
B.P. 138, 21004 Dijon Cedex, France
\end{center}

%\maketitle
\begin{abstract}
We discuss the relationship between Schlesinger system and stationary
axisymmetric Einstein's equation on the level of algebro-geometric
solutions. In particular, we calculate all metric coefficients corresponding
to solutions of Ernst equation in terms of theta-functions constructed in 
\cite{Koro88,KorMat89,Koro96}.

{\bf Mathematics Subject Classification (1991):}
\vspace{24pt}\\
 
\end{abstract}

\newpage
\section{Introduction}
\setcounter{equation}{0}

Interaction between algebraic geometry of the compact Riemann surfaces and the theory 
of integrable systems represent nowadays a well established paradigma of modern
mathematical physics. The analysis on the compact Riemann  surfaces was essentially completed in 
19th century in the famous classical works by Gauss, Euler, Riemann, Jacobi, Weierstrass
etc. The penetration of the tools provided by their methods in the theory of integrable systems 
started in the same time. Legrange first solved the equations of motion of the Euler top
by means of the use of the Jacobi elliptic functions. Next essential breakthrough was achieved 
in 1889 by
Kovalevski \cite{Kova889} who found the new integrable case of the motion of rigid body
solvable in terms of genus 2 algebraic functions. Slightly later Dobriner published 
the explicit solution of sine-Gordon equation  expressed in terms of two-dimensional
theta-functions \cite{Dobr}. Then Carl Neumann solved the equations of geodesic motion 
on three-axis ellipsoid in terms of two-dimensional hyperelliptic theta-functions
\cite{Neum}.

The same time it started the development of the spectral theory of the Sturm-Liouville
operators with periodic coefficients.  First important steps in the field were done by
Floke \cite{Flok}, Lyapunov  \cite{Lyap} and others (see related references in the book
\cite{Glaz}.

In 1919th french matematician Jule Drach wrote the remarkable article  (absolutely forgotten
for next 60 years) devoted to the construction of explicitly solvable Sturm-Liouville 
equations associated to hyperelliptic curves of any genus \cite{Drac19}.

One of the subsequent developments of the discovery of the inverse scattering transform (IST) method by 
Gardner, Green, Kruskal and Miura \cite{GGKM66} was the complete understanding of the deep
interplay between the classical works listed above. This connection now became a part of what is
called algebrogeometric approach to the solution of nonlinear differential equations.
This field of activity was initiated in 1974-1976 by the works of (in chronological order)
Novikov, Lax, M.Kac, Its, Matveev, Dubrovin, McKean, van Moerbeke and Krichever.

Speaking about the aspects of the theory connected to the explicit solutions of these 
equations one should mention the formulas for solutions of the Korteveg-deVries equation 
with periodic initial data obtained in \cite{ItsMat75}. Soon after the same kind of 
formulas were derived for numerous other equations integrable via IST method, including
Non-linear Schr\"{o}dinger \cite{NS}, Sine-Gordon   \cite{SG}, Toda lattice  \cite{TL},
Kadomtzev-Petviashvili \cite{KP} and so on. The generic feature of the algebro-geometric 
solutions of the systems listed above is that they are parametrised by the fixed algebraic curves
and the associated dynamics is linear on their Jacobians. These results found many 
beautiful and unexpected applications in various branches of modern mathematics and physics,
including differential geometry of surfaces \cite{Bobe93} and algebraic geometry
(Novikov hypothesis \cite{Shiota}). 

It is also relevant to notice that simultaneously with the soliton theory, the 
algebro-geometric methods were actively applied in the framework of twistor theory
(which, essentially, also deals with integrable equations, and exploits 
the methods very closely related to ones used by integrable systems  community, but essentially focuses
on the global properties of the solutions). Probably, the main
achievments in this direction were classifications of the instanton and self-dual
monopole configurations (see for more details the book \cite{AtiHit} and references therein).

Technically and conceptually new development was the application of algebrogeometric ideology
to the Einstein equations of general relativity in presence of two commuting Killing vectors
(Ernst equation). The embedding of this equation in the framework of the IST
approach was initiated by Belinskii-Zakharov \cite{BelZak78} and Maison \cite{Mais78}
\footnote{Coming back to the works of classics of differential geometry of 19th century 
it is curious to mention that essentially the same Lax pair appeared in the
work of Bianchi \cite{Bian??} in the study of so-called Bianchi congruences.}.
The characteristic feature of the associated zero-curvature representation is the
non-trivial dependence of the associated connection on the spectral parameter. Namely, the 
connection lives on the genus zero algebraic curve depending on space-time variables.
This peculiarity entails the drastic change of the construction and qualitative 
properties of algebro-geometric solutions first obtained in 1988 \cite{Koro88,KorMat89}.
Dynamics in these solutions is generated by hyperelliptic curves with two coordinate-dependent 
branch points. The algebro-geometric solutions of the Ernst equation do not possess any 
periodicity properties which were inevitable for all KdV-like cases studied before.
Moreover, we can explicitly incorporate in the construction an arbitrary functional 
parameter which never appears in the traditional KdV-like 1+1 integrable systems.
Wide subclass of the obtained solutions turns out to be asymptotically flat \cite{Koro93}.
As a simple degenerate case the algebro-geometric solutions contain the whole
class of multisoliton solutions found by Belinskii and Zakharov \cite{BelZak78}.

Simplest elliptic solutions  were studied in  \cite{Koro93}. Despite many interesting properties,
they contain ring-like naked singularities making it difficult to exploit them in a real
physical context. More realistic physical application of the particular family of
the class of algebro-geometric solutions came out from the series of papers of
Meinel, Neugebauer and their collaborators starting from 1993 
\cite{MeiNeu94, MeiNeu95}.
These works were devoted to the investigation of the boundary-value problem corresponding to the
infinitely thin rigid relativistically-rotating dust disc. The explicit embedding of the
dust disc solution in the formulas of \cite{Koro88} was given  in \cite{Koro96}.
In the subsequent series of papers Klein and Richter (see, for example, \cite{KleRic96,KleRic97}) established
the link of the algebrogeometric solutions of Ernst equation given in \cite{Koro88,Koro96}
with the scalar Riemann-Hilbert problem on hyperelliptic curve and further discussed 
potential applications to rotating bodies.
 
Purpose of this paper is twofold. First, we 
extend the link between isomonodromic solutions of the Ernst equation
and classical Schlesinger system \cite{Schl12}, established in the paper \cite{KorNic95}, on
the level of algebro-geometric solutions. Namely, we show how all algebro-geometric
solutions of the Ernst equation may be obtained from the algebro-geometric solutions of 
the Schlesinger system found in the recent paper \cite{KitKor98}.
This allowes to get remarkably  short expression for  algebro-geometric solutions of the
Ernst equation:
\be
\E(\xi,\xb)=
\frac{\Th\left[^\rb_\sb\right]\left(V\big|_\xi^{\infty^1}\right)}
{\Th\left[^\rb_\sb\right]\left(V\big|_\xi^{\infty^2}\right)}\;,
\la{solint}\ee
in terms of theta-functions with constant characteristics $\rb,\sb\in\R^{g_0}$,
associated to hyperelliptic curve $\L_0$ of genus $g_0$ with two 
coordinate-dependent branch points:
\be
\nu^2=(w-\xi)(w-\xb)\prod_{j=1}^{2g_0} (w-w_j)\;.
\la{L0intr}\ee
Applying certain limiting procedure to this solution, we can get general
algebro-geometric solution  of the Ernst equation.

Second, we give explicit expressions 
for all metric coefficients corresponding to algebro-geometric solutions of 
the Ernst equation. The most non-trivial part is the calculation of so-called 
conformal factor; for this we use the link between the conformal factor 
and tau-function of the Schlesinger system, established in \cite{KorNic95},
and the formula for the tau-function obtained in \cite{KitKor98}. The final formula
for the conformal factor corresponding to solution (\ref{solint}) looks as follows: 
\be
e^{2k}= 
\f{\Th\left[^\pb_\qb\right]\left(0 |\B\right)}
{\sqrt{\det {\cal A}_0}}\prod_{j=1}^{2g_0}|w_j-\xi|^{-1/4}\;,
\la{confacintr}\ee
where  ${\cal A}_0$ - is the matrix of  $a$-periods of holomrphic differentials 
$w^j dw/\nu\;,\;\; j=1,\dots, g_0$ on $\L_0$, and the theta-function is associated to the
hyperelliptic curve $\L$ of genus $2g_0-1$ defined by the equation
\ben
y^2=\prod_{j=1}^{2g+2}(\g-\g_j)\;,
\een
where 
\ben
\g_j=\f{2}{\xi-\xb}\Big\{w-\f{\xi+\xb}{2}+\sqrt{(w-\xi)(w-\xb)}\Big\}\;.
\een
Appropriate limiting procedure allows to deduce from this formula the expression for 
conformal factor corresponding to general algebro-geometric solutions.
Notice, that it seems to be impossible to give simple expression for the conformal factor 
entirely in terms of the objects corresponding to curve $\L_0$, which was the obstacle
for deriving this formula without understanding the link between the conformal factor and
the tau-function.

\section{Schlesinger system and stationary axisymmetric Einstein equations}
\setcounter{equation}{0}
\subsection{Schlesinger system}

Consider the following linear differential equation for function
$\Psi(\lambda)\in SL(2,\C)$:
\be
\f{d\Psi}{d\g}= A(\g)\Psi
\la{ls}\ee
where 
\be
A(\g)= \sum_{j=1}^N\f{A_j}{\g-\g_j}
\la{A}\ee
and matrices $A_j\in sl(2,\C)$ are independent of $\g$.
Let us impose the initial condition
\be
\Psi(\g=\infty)=I  
\la{norm}\ee
Function $\Psi(\g)$ defined by (\ref{ls}) and (\ref{norm}) lives on the
universal covering $X$ of $\C P^1\setminus\{\g_1,\dots,\g_N\}$. The 
asymptotical expansion of $\Psi(\g)$ near singularities $\g_j$ is
given by
\be
\Psi(\g)= Q_j(I+ O(\g-\g_j)) (\g-\g_j)^{T_j} C_j
\la{asymp}\ee
where $Q_j,\;C_j\;\in SL(2,\C)$ and   $T_j$ is traceless diagonal matrix.
Matrices 
\be
M_j=C_j^{-1} e^{2\pi i T_j} C_j, \hskip1.0cm j=1,\dots, N 
\la{Mj}\ee
are called monodromy matrices.

The assumtion of independence of all matrices $M_j$ of the parameters $\g_j$:
\be
\f{\p M_j}{\p\g_k}=0
\la{iso}\ee
is called the isomonodromy condition; it implies the following 
dependence of $\Psi(\g)$ on $\g_j$, which can be deduced from
(\ref{asymp}):
\be
\f{\p\Psi}{\p\g_j}=-\f{A_j}{\g-\g_j}\Psi
\la{ls1}\ee
The compatibility condition of (\ref{ls}) and (\ref{ls1}) is equivalent to the
Schlesinger system \c{Schl12} for the residues $A_j$:
\be
\frac{\partial A_j}{\partial\g_i}=
\frac{[A_i,A_j]}{\g_i-\g_j},\;\;\;i\neq j\;,\;\;\;\;\;
\frac{\partial A_i}{\partial\g_i}=
-\sum_{j\neq i}
\frac{[A_i,A_j]}{\g_i-\g_j}.
\label{sch} 
\ee

The Schlesinger system admits the ``multi-time'' Hamiltonian formulation
(\cite{JimMiw81}) with respect to the following Poisson structure:
\be
\{A_j^a, A_k^b\}= f^{ab}_c A_k^c \delta_{jk}
\la{PS}\ee
where $f^{ab}_c$ are structure constants of $sl(2)$. Evolution with respect 
to ``times'' $\g_j$ is described by the Hamiltonians
\be
H_j=\sum_{k\neq j}\f{\tr A_j A_k}{\g_j-\g_k}
\la{Hj}\ee    
The function $\tau (\{\g_j\})$, generating Hamiltonians $H_j$ according to
equations 
\be
\f{\p}{\p\g_j}\log\tau = H_j
\la{tauHj}
\ee
is called the $\tau$-function of Schlesinger system. Compatibility of 
equations (\ref{tauHj}) follows from the Poisson commutativity of all
Hamiltonians $H_j$.

\subsection{Stationary axisymmetric Einstein equations from Schlesinger 
system}

The Einstein equations for the line element 
\be
ds^2=f^{-1}[e^{2k}(dz^2+d\rho^2)+\rho^2 d\phi^2]-f (dt+ F d\phi)^2
\la{metric}\ee
where all metric coefficients $f,k,F$ are assumed to depend only on
$\rho$ and $z$, reduce  to the Ernst equation
\be
(\E+\Eb)(\E_{zz}+\f{1}{\rho}\E_\rho+\E_{\rho\rho})=2(\E_z^2+\E_\rho^2).
\la{EE}\ee
The metric coefficients may be restored from the complex-valued Ernst potential 
$\E(z,\rho)$ according to the following equations:
\be
f=\Re\E 
\hskip1.0cm 
F_\xi=2\rho\f{(\E-\Eb)_\xi}{(\E+\Eb)^2}
\hskip1.0cm
k_\xi=2i\rho\f{\E_\xi\Eb_\xi}{(\E+\Eb)^2}
\la{coeff}\ee
where $\xi = z + i\rho$.

Equation (\ref{EE}) is the compatibility condition of the following
linear system:
\be
\Psi_\xi =\f{G_\xi G^{-1}}{1-\g}\Psi\hskip1.0cm
\Psi_\xb =\f{G_\xb G^{-1}}{1+\g}\Psi
\la{lsg}\ee
where
\be
\g=\f{2}{\xi-\xb}\left\{\l-\f{\xi+\xb}{2}+\sqrt{(\l-\xi)(\l-\xb)}\right\},
\la{gamma}\ee
$\l\in \C$ is a spectral parameter and
\be
G=\f{1}{\E+\Eb}\left(\ba{cc} 2 & i(\E-\Eb)\\
                            i(\E-\Eb) & 2\E\Eb\ea\right)
\la{g}\ee
In terms of the matrix $G$ the Ernst equation may be equivalently rewritten
as follows:
\be
(\rho G_\rho G^{-1})_\rho + (\rho G_z G^{-1})_z = 0
\la{EEg}\ee

Relationship between solutions of Schlesinger system and Ernst equation was 
revealed in  \cite{KorNic95}:
\begin{theorem}
Let $\{A_j\}$ be some solution of the Schlesinger system (\ref{sch})
and $\Psi(\g)$ be related solution of equation (\ref{ls}) satisfying the
following conditions:
\be
\Psi^t(\f{1}{\g})\Psi(0)^{-1}\Psi(\g) = I\;,
\la{coset}\ee
\be
\Psi(-\bar{\g})= \overline{\Psi(\g)}\;.  
\la{reality}\ee
Let in addition  $\g_j=\g(\l_j,\xi,\xb),\;\l_j\in \C$ for all $j$, 
with  function $\g(\l,\xi,\xb)$  given by (\ref{gamma}).   
Then 
\be
G(\xi,\xb)\equiv \Psi(\g=0,\xi,\xb)
\la{sol1}\ee
is a solution of Ernst equation (\ref{EEg}). Function 
$\Psi(\g,\{\g_j\})$ solves the associated linear system (\ref{lsg}).
Metric coefficient $e^{2k}$ of the line element (\ref{metric}) 
is related to the $\tau$-function of the Schlesinger system as 
follows:
\be
e^{2k}=C \prod_{j=1}^N \left\{\f{\p\g_j}{\p \l_j}\right\}^{\tr A_j^2/2} \tau\;,
\la{taucon}\ee
where $C$ is a constant of integration.
\la{ErnstSchl}
\end{theorem}

\section{Solutions of the Schlesinger system in terms of theta-functions}
\setcounter{equation}{0}
Let $N=2\g+2$. Define hyperelliptic curve $\L$ of genus $\g$ by the equation
\be
w^2=\prod_{j=1}^{2g+2}(\g-\g_j)
\la{L}\ee 
with 
 basic cycles $(a_j,b_j)$ chosen according to 
figure~ 1.
%\ref{KORCICYCLE1}.
%\begin{figure}[htb]
%\begin{center}
%%\input{KORCICYCLE1.pstex_t}
%%\end{center}
%\caption{Branch cuts and canonical basis of cycles on the 
%hyperelliptic curve, $\L$.
%Continuous and dashed paths lie on the first and the second sheet of $\L$,
%respectively.}
%\label{KORCICYCLE1}
%\end{center}
%\end{figure}

%Let us denote the fundamental polygon of $\L$ by $\hat{\L}$.
The basic  holomorphic 1-forms on $\L$ 
are given by
\be
\f{\g^{k-1}d\g}{w},\;\;\;\;\;\;k=1,\dots,g.
\la{dUk0}\ee
Let us define $g\times g$ matrices of $a$- and $b$-periods 
of these 1-forms by
\be
{\cal A}_{kj}=\oint_{a_j}\f{\g^{k-1}d\g}{w},\;\;\;\;\;\;\;
{\cal B}_{kj}=\oint_{b_j}\f{\g^{k-1}d\g}{w}.
\la{AB}\ee
Then the holomorphic 1-forms 
\be
dU_k=\f{1}{w}\sum_{j=1}^g ({\cal A}^{-1})_{kj} \g^{j-1} d \g
\la{dUk}\ee
satisfy the normalization conditions
$\oint_{a_j} dU_k=\delta_{jk}$. 

The matrices ${\cal A}$ and ${\cal B}$ define the symmetric 
$g\times g$ matrix of $b$-periods of the curve $\L$:
$$\B= {\cal A}^{-1}{\cal B}\;.$$ 
Let us now introduce the theta function with characteristic $[^\pb_\qb]$ ($\pb\in\C^g$, $\qb\in\C^g$) by the following series, 
\be
\Th[^\pb_\qb](\z|\B)=\sum_{\m\in \Z^g} \exp\{\pi i \langle\B(\m+\pb),\m+\pb\rangle +2\pi i \langle\z+\qb,\m+\pb\rangle\},
\la{theta}\ee
for any $\z\in\C^g$. It possesses 
the following periodicity properties:
\be
\Th[^\pb_\qb](\z+\eb_j)=
e^{2\pi i p_j} \Th[^\pb_\qb](\z),
\la{pera}\ee
\be
\Th\left[^\pb_\qb\right](\z+\B \eb_j)=e^{-2\pi i q_j}e^{-\pi i \B_{jj}-2\pi i \z_j}
\Th[^\pb_\qb](\z),
\la{perb}\ee
where
\be
\eb_j\equiv (0,\dots,1,\dots,0)^t
\ee
($1$ stands in the $j$th place). 

The theta-function with characteristics is related as follows to the
theta-function without characteristics:
\be
\Th[^\pb_\qb](\z)= \Th(\z+\B\pb+\qb) e^{\pi i \langle\B\pb,\pb\rangle+
2\pi i \langle\pb,\z+\qb\rangle}
\la{charac}
\ee

Cut curve $\L$ along all basic cycles to get the fundamental polygon   $\Lh$. 
For any meromorphic 1-form $dW$ on $\L$ we shall use the notation 
\ben
W\big|_Q^P\equiv\int_Q^P dW
\een
where the integration contour lies inside of $\Lh$ (if $dW$ is meromorphic,
the value of this integral might also depend on the choice of integration
contour inside of $\Lh$).
By $U\big|_Q^P$ we shall denote vector with components $U_j|_Q^P$.
The vector of Riemann constants corresponding to our choice of the 
initial point of this map reads as follows \cite{Fay}:
\be
K=\f{1}{2}\B(\eb_1+\dots+\eb_g) +\f{1}{2}(\eb_1+2\eb_2\dots+g\eb_g).       
\la{K}\ee

The characteristic with components $\pb\in\C^g/2\C^g$,
$\qb\in\C^g/2\C^g$ is called half-integer characteristic: the half-integer 
 characteristics are in one-to-one correspondence with the half-periods
$\B\pb+\qb$.
If the scalar product 
$4\langle\pb,\qb\rangle$ is odd, then the related theta function is 
odd with respect to its argument $\z$ and the characteristic $[^\pb_\qb]$ is 
called odd, and if this scalar product is even, then the theta function 
$\Th[^\pb_\qb](\z)$ is even with respect to $\z$ and the characteristic 
$[^\pb_\qb]$ is called even.

The odd characteristics which will be of importance for us in the sequel correspond to any given subset $S=\{\g_{i_1},\dots,\g_{i_{g-1}}\}$ of $g-1$ arbitrary non-coinciding branch points. The odd half-period associated to the subset $S$ is given by  
\be
\B\pb^S+\qb^S= \sum_{j=1}^{g-1}U\Big|_{\g_{1}}^{\g_{i_j}}   - K.
\la{odd}
\ee
where $dU=(dU_1,\dots,dU_g)^t$.
Denote by 
 $\Omega_\g\subset\C$ the neighbourhood of the infinite point $\g=\infty$,
such that $\Omega_\g$ does not overlap with projections of all basic cycles
on $\g$-plane.
Let the $2\times 2$ matrix-valued function $\Phi(\g)$ be defined in the 
domain $\O_\g$ of the first sheet of $\L$  by the following formula,
\be
\Phi(\g\in\O_\g)=\left(\ba{cc}\phi(\g)\;\;\;\;\; \phi(\g^*)\\
                  \psi(\g)\;\;\;\;\; \psi(\g^*)\ea\right),
\la{Phi}\ee 
where functions $\phi$ and  $\psi$ are defined in the fundamental
polygon $\Lh$ by the formulas:
\be
\phi(\g)=\Th\left[^\pb_\qb\right]\left(U\big|_{\g_1}^\g + U\big|_{\g_1}^{\g_\phi}\Big|\B\right)\Th\left[S\right]\left(U\big|_{\g_\phi}^\g 
\Big|\B\right),
\la{phi1}
\ee
\be
\psi(\g)=\Th\left[^\pb_\qb\right]\left(U\big|_{\g_1}^\g + U\big|_{\g_1}^{\g_\psi}\Big|\B\right)\Th\left[S\right]\left(U\big|_{\g_\psi}^\g 
\Big|\B\right),
\la{psi1}
\ee
with two arbitrary (possibly $\{\g_j\}$-dependent) points $\g_{\phi}$, 
$\g_\psi\in\L$ and arbitrary constant characteristic $\left[^\pb_\qb\right]$; $*$ is the involution on $\L$ interchanging the sheets;
\ben
\Th\left[S\right]\left({\bf z}\right)
\equiv\Th\left[^{\pb^S}_{\qb^S}\right]\left({\bf z}\right)
\een
where odd theta characteristic $\left[^{\pb^\0}_{\qb^\0}\right]$ corresponds to an arbitrary subset $S$ of $g-1$ branch points via  {\rm Eq.~(\ref{odd})}.

Since domain $\O_\g$ does not overlap with projections of all basic cycles
of $\L$ on $\g$-plane, domain $\L_\g^*$ does not overlap with the boundary
of $\Lh$, and functions $\phi(\g^*)$ and $\psi(\g^*)$ in (\ref{Phi}) are
uniquely defined by (\ref{phi1}), (\ref{psi1}) for $\g\in\O_\g$.

Now choose some sheet of the universal covering $X$, 
define new function $\Psi(\g)$ in subset $\Omega_\g$ of this sheet 
by the formula 
\be
\Psi(\g\in\Omega_\g)=\sqrt{\f{\det \Phi(\infty^1)}{\det \Phi(\g)}}\Phi^{-1}(\infty^1)\Phi(\g)
\la{Psi}\ee
and extend on the rest of  $X$ by analytical continuation.

Function  $\Psi(\g)$ (\ref{Psi}) transforms as follows with respect
to the tracing around basic cycles of $\L$ (by $T_{a_j}$ and $T_{b_j}$ 
we denote corresponding transport operators):
\ben
T_{a_j}[\Psi(\g)]=\Psi(\g) M_{a_j}\;; \hskip1.0cm
T_{b_j}[\Psi(\g)]=\Psi(\g) M_{b_j}\;,\hskip1.0cm j=1,\dots,2g_0-1\;,
\een
where 
\be
 M_{a_j}=\left(\ba{cc} e^{2\pi i p_j} & 0 \\
                         0 &  e^{-2\pi i p_j}\ea\right)\;,\hskip1.0cm
 M_{b_j}=\left(\ba{cc} e^{-2\pi i q_j} & 0 \\
                         0 &  e^{2\pi i q_j}\ea\right)\;.
\la{MaMb}
\ee

The following statement was proved in the paper \cite{KitKor98}:

\begin{theorem}\la{theoPsi}
Let $\pb,\qb\in \C^g$ be an arbitrary set of $2g$ constants such that
$\left[^\pb_\qb\right]$ is not half-integer characteristic. Then:
\begin{enumerate}
\item
Function $\Psi(Q\in X)$ defined by (\ref{Psi}) is independent of $\g_\phi$
and $\g_\psi$ and satisfies the linear system (\ref{ls}) with 
\be
A_j\equiv {\rm res}|_{\g=\g_j} \left\{\Psi_\g\Psi^{-1}\right\}.
\la{Aj}\ee
which in turn solve the Schlesinger system (\ref{sch}).
\item
Monodromies (\ref{Mj}) of $\Psi(\g)$ around points $\g_j$ are given by
\be
M_j= \left(\ba{cc} 0 & -m_j \\m_j^{-1} & 0 \ea\right)\;,
\la{Mj1}\ee
where constants $m_j$ may be expressed in terms of $\pb$ and $\qb$
(see \cite{KitKor98}).
\item
The $\tau$-function, corresponding to solution  (\ref{Aj}) of the 
Schlesinger system, has the following form:
\be
\tau(\{\g_j\})=[\det{\cal A}]^{-\f 12}
\prod\limits_{j<k}(\g_j-\g_k)^{-\frac 18}\Theta\left[^\pb_\qb\right](0|\B)
\la{tau}\ee
\end{enumerate}
\end{theorem}

\section{Solutions of the Ernst equation in terms of theta-functions.
Formulas for the metric coefficients}
\setcounter{equation}{0}
According to the relationship between Schlesinger system and Ernst equation
given by the theorem \ref{ErnstSchl}, we can derive solutions of Ernst
equation in terms of theta-functions from construction of the theorem 
\ref{theoPsi}.  The necessary additional work to do is to 
choose the parameters of the construction (i.e. the constants
$\l_j$ and vectors $\pb,\;\qb$) to provide the constraints
(\ref{coset}) and (\ref{reality}). To get these constraints 
fulfilled we have to assume that the 
curve $\L$ is invariant under the holomorphic involution 
$\sigma$ acting on every sheet of $\L$ as
\be
\sigma\;:\; \g\rightarrow \f{1}{\g}\;,
\la{sigma}
\ee
and anti-holomorphic involution $\mu$ acting on every sheet of $\L$ as
\be
\mu\;:\;\g\to -\bar{\g}\;.
\la{mui}\ee
Constraints (\ref{coset}) and (\ref{reality}) turn out to be 
compatible with each other only if the genus $g$ is odd:
\be
g = 2g_0 - 1 \;.
\la{gg0}\ee
We shall enumerate the branch points $\g_j,\;j=1,\dots, 4 g_0$ in such an 
order that
\be
\g_j=\g_{j+2g_0}^{-1}\;,\hskip0.5cm j=1,\dots,2g_0
\la{invbp}\ee
and for some $k\leq g_0$
\ben
\g_{j}\in i\R\;,\hskip0.5cm 1\leq j\leq 2k\;;\hskip0.8cm
\g_{2j+1}+\bar{\g}_{2j+2} =0\;,\;\;\;2k+1\leq j\leq 2 g_0-1 .
%\la{rcg}
\een
We shall now distinquish two cases:
\begin{enumerate}
\rm
\item
Curve $\L$ is non-separable
\footnote{Curve admitting anti-holomorphic involution is called separable,
if it is divided into two pieces by the ovals invariant with respect to 
the anti-involution, and non-separable otherwise.} 
 with respect to the action of anti-involution 
$\mu$, i.e. $k\geq 1$. Then the basic cycles  
$(a_j,b_j)\;,\;\;j=1,\dots, 2g_0-1$
on $\L$ may be chosen as shown in figure 2a.
\item 
Curve $\L$ is separable with respect to the action of anti-involution 
$\mu$, i.e. $k=0$ and, therefore,  none of the points $\g_j$ lie on the imaginary axis.
In this case we choose  the basic cycles  on $\L$ as shown in figure 2b.

\end{enumerate}

In both cases the basic cycles transform in the 
following way under the action of the involution $\sigma$:
\be
\s(a_1) = -a_1\hskip1.0cm \s(b_1) = -b_1
\la{ab0s}\ee
\be
\s(a_j) = a_{j+g_0-1}\hskip1.0cm \s(b_j) = b_{j+g_0-1}\;,\hskip1.0cm
2\leq j\leq g_0
\la{abjs}\ee

Constraint (\ref{coset}) leads to  the following equations for
$M_{a_j}$ and $M_{b_j}$ (\ref{MaMb}):
\ben
M_{a_j}^t M_{a_{j+g_0-1}}= I \;;\hskip1.0cm M_{b_j}^t M_{b_{j+g_0-1}}= I\;,
 \hskip1.0cm 2\leq j\leq  g_0\;.
\een
(Equations $M_{a_1}^t= M_{a_1}$ and  $M_{b_1}^t= M_{b_1}$, which arise from
the calculation of monodromies of (\ref{coset}) along the basic cycles
$a_1$ and $b_1$, are  identically fulfilled) 
 
In turn, we get for
components of the  vectors $\pb$ and $\qb$ the following equations
\be
p_j+ p_{j+g_0-1}=0\;;\hskip1.0cm q_j+ q_{j+g_0-1}=0\;,\hskip1.0cm 
2\leq j\leq g_0
\la{pqs}\ee

It remains to derive constraints on $\pb$ and $\qb$ imposed by  
the ``reality conditions'' (\ref{reality}).
We  shall consider two cases separately.
\begin{enumerate}\rm
\item
{\it Non-separable case ($k\geq 1$).} 
The basic cycles of $\L$ shown in figure 2a 
behave as follows with respect to the action of the anti-involution $\mu$:
\ben
\mu(a_j) = -a_j\;,\hskip0.5cm\forall j\;;\hskip1.0cm
\mu(b_1) = b_1+2 a_1
\een
\ben
\mu(b_j) = b_j\;,\hskip0.5cm 2\leq j\leq k\;;\hskip1.0cm
\mu(b_j) = b_j-a_j\;,\hskip0.5cm k+1\leq j\leq g_0
\een
(since we assumed already the invariance of $\L$ under the 
involution $\s$, it is sufficient to determine the  action of $\mu$ only
on the first $g_0$ cycles $b_j$).
Now from (\ref{reality})  we get the
following equations for monodromies:
\ben
\ol{M}_{a_j}= M_{a_j}^{-1}\;,\hskip0.5cm \forall j\;;\hskip1.0cm
\ol{M}_{b_1}= M_{b_1} M^2_{a_1}\;; 
\een
\ben 
\ol{M}_{b_j}=M_{b_j}\;,\hskip0.5cm  2\leq j\leq k  \;;\hskip1.0cm
\ol{M}_{b_j}=M_{b_j} M_{a_j}^{-1}\;,\hskip0.5cm k+1\leq j\leq g_0\;,
\een
which is equivalent to the following conditions imposed on 
$\pb$ and $\qb$:
\ben
p_j\in\R\;,\hskip0.5cm \;\forall j;\hskip1.0cm
\Re q_1=p_1\;;
\een
\be
\Re q_j=0\;,\hskip0.5cm  2\leq j\leq k\;;\hskip1.0cm
\Re q_j=-\f{1}{2}p_j\;,\hskip0.5cm  k+1\leq j\leq g_0 \;.
\la{pqns}\ee

\item
{\it Separable case ($k=0$).} 
In this case we shall choose the basic cycles according to figure 2b. 
They transforms under the action of $\mu$ in the following way:
\ben
\mu(a_j) = -a_j\;,\hskip0.5cm\forall j \;;\hskip1.0cm
\mu(b_1) = b_1+\sum_{l=2}^{g_0} (a_{l+g_0-1} -a_l)\;;
\een
\ben
\mu(b_j) = b_j- a_j+\sum_{l=1}^{g_0} a_l\;,\hskip0.5cm 2\leq j\leq g_0\;, 
\een
which leads to the following conditions on $\pb$ and $\qb$:
\be
p_j\in\R\;,\hskip0.5cm \;\forall j\;;\hskip1.0cm
\Re q_1=-\f{1}{2}\sum_{l=2}^{g_0} p_l\;;\hskip1.0cm
\Re q_j=-\f{1}{2}\sum_{l=1}^{g_0} p_l-\f{p_j}{2}\;.
\la{pqsep}
\ee

\end{enumerate}

Now, taking into account the theorems \ref{ErnstSchl} and \ref{theoPsi},
we get the following 

\begin{theorem}
\la{approx}
Let the genus of curve $\L$ be odd: $g=2g_0-1$; and the basic cycles be 
chosen according to
figure 2a if $\L$ is of non-separable and figure 2b  if $\L$ is 
of separable  type.  
Let  $\pb,\qb\in \C^{2g_0-1}$ be arbitrary constant vectors 
satisfying conditions (\ref{pqs}). Let in addition $\pb,\qb$ 
satisfy conditions (\ref{pqns}) if $\L$ is non-separable and (\ref{pqsep})
if $\L$ is separable.
Define function $\Psi(\g)$ by the expressions (\ref{Phi}),
(\ref{phi1}),(\ref{psi1}) and (\ref{Psi}). Then
the function
\be
G(x,\rho)\equiv\Psi(x,\rho,\g=0) 
\la{sol2}\ee
 satisfies the 
Ernst equation (\ref{EEg}) and may be represented in the form (\ref{g}).
\end{theorem}

The expression for the Ernst potential $\E$ may be obtained from
(\ref{sol2}) in terms of theta-functions associated to curve $\L$. 
However, it may be essentially simplified if we make use of 
invariance of curve $\L$ under involution $\s$ and use the spectral 
parameter $\l$ from (\ref{gamma}) instead of $\g$. Namely, curve $\L$ may
be represented as fourfold covering of $\l$-plane; its Hurvitz
diagram is shown in figure 3. In this realization 
involution $\s$ interchanges the sheets $1\leftrightarrow 3$ and
$2\leftrightarrow 4$; involution $*$ interchanges the sheets 
$1\leftrightarrow 2$ and $3\leftrightarrow 4$.

Now introduce new hyperelliptic curve $\L_0$ of genus $g_0$ 
defined by   equation
\be
\nu^2=(\l-\xi)(\l-\xb)\prod_{j=1}^{2g_0} (\l-\l_j)
\la{L0}\ee
Curve $\L_0$ together with canonical basis of cycles $(a_j^0,b_j^0)$
is shown in figure 4. 
Introduce the  dual basis $dV=(dV_1,\dots,dV_{g_0})^t$ of holomorphic 
1-forms on $\L_0$  by
\be
dV_j=\f{1}{\nu}\sum_{k=1}^{g_0} ({\cal A}_0^{-1})_{jk}\l^{k-1} d\l\;,\hskip0.5cm 
j=1,\dots, g_0\;,
\la{dVj}\ee
where 
\be
({\cal A}_0)_{kj}\equiv \oint_{a_j^0} \f{\l^{k-1} d\l}{\nu}\;,
\hskip0.5cm j,k=1,\dots, g_0\;,
\la{A0}\ee
and corresponding matrix of $b$-periods $\B_0$.

Curve $\L$ is twofold non-ramified covering  of $\L_0$: $\Pi\;:\L\to\L_0$,
such that the points of $\L$ related by the involution $\sigma$ project onto
the same point of $\L_0$, namely,
the $\l$-sheets $1$ and $3$ of $\L$ are projected onto the
1st sheet of $\L_0$, and sheets $2$ and $4$ of $\L$ are projected onto
the 2nd sheet of $\L_0$, preserving projections of corresponding 
points on $\l$-plane. Anti-involution $\mu$ inherited from $\L$ acts
on every sheet of $\L_0$ as $\l\rightarrow\bar{\l}$.

Existence of reduction (\ref{coset}) allows to 
alternatively express function $\Psi$ 
 in terms of theta-functions associated to the curve $\L_0$.
Denote by $\O_\l$ the  neighbourhood of the point $\l=\infty$ 
being the projection of the domain $\O_\g$ into  $\l$-plane.
Define function $\Phi_0(\l\in\O_\l)$ in the domain $\O_\l$ lying 
on the first sheet of $\L_0$ by the following formula:
\be
\Phi_0(\l\in\O_\l)=\left(\ba{cc}\phi_0(\l)\;\;\;\;\; \phi_0(\l^*)\\
                  \psi_0(\l)\;\;\;\;\; \psi_0(\l^*)\ea\right),
\la{Phi0}\ee 
where involution $*$ inherited on $\L_0$ from $\L$ interchanges the
$\l$-sheets of $\L_0$;
\be
\phi_0(\l) = \Th\left[^\rb_\sb\right]\left(V\big|_\xi^\l \Big|\B_0\right)\; ,
\hskip1.0cm
\psi_0(\l) = \Th\left[^\rb_\sb\right]\left(V\big|_\xb^\l \Big|\B_0\right)\; ,
\la{phipsi0}\ee
\be
\phi_0(\l^*) = -i\Th\left[^\rb_\sb\right]\left(-V\big|_\xi^\l \Big|\B_0\right)\; ,
\hskip1.0cm
\psi_0(\l^*) = i\Th\left[^\rb_\sb\right]\left(-V\big|_\xb^\l dV\Big|\B_0\right)\; ,
\la{phipsi01}\ee
for $\l\in\O_\l$,
and constant vectors $\rb,\;\sb\in \C^{g_0}$ satisfy the following
reality conditions:
\be
\rb\in \R^{g_0}
\la{abj}\ee
\be
\Re\sb_j = \sum_{l=1}^{g_0}\f{\rb_l}{2}\;,\hskip0.5cm  1\leq j\leq k\;;
\hskip1.0cm  \Re\sb_j = \sum_{l=1,\;l\neq j}^{g_0}\f{\rb_l}{2}\;
,\hskip0.5cm k+1 \leq j\leq g_0 \;.
\la{bbj}\ee  

The basic cycles of curve $\L_0$ behave as follows under the action of
anti-involution $\mu$:
\ben
\mu(a_j^0)=-a_j^0\;,\hskip0.5cm\forall j \; ;\hskip1.0cm
\mu(b_j^0)=b_j^0+\sum_{l=1}^{g_0} a_l^0\;,\hskip0.5cm j\leq k\; ;\hskip1.0cm 
\mu(b_j^0)=b_j^0+\sum_{l\neq j} a_l^0\;,\hskip0.5cm j > k
\een
which implies the following relations for the matrix of $b$-periods 
of $\L_0$:
\ben
\Re(\B_0)_{jl}= - \f{1}{2}\;,\hskip0.5cm j\leq k \; ;\hskip1.0cm
\Re(\B_0)_{jl}= - \f{1}{2} +\f{\delta_{jl}}{2}\;,\hskip0.5cm j > k
\een
Now the relations (\ref{abj}), (\ref{bbj}) may be equivalently represented as
\be
\Re(\B_0\rb+\sb) = 0\;;
\la{rcalt}\ee
this, in particular, implies the following relation between functions
$\phi_0$ and $\psi_0$:
\be
\psi_0(P)=\overline{\phi_0 (\bar{P})}
\la{phi0psi0}\ee

Now define function 
\be
\Psi_0(\l\in \O_\l)\equiv \sqrt{\f{\det \Phi_0(\infty^1)}{\det \Phi_0(\l)}}\Phi_0^{-1}(\infty^1)\Phi_0(\l)\;,
\la{Psi00}
\ee
and extend it to the universal covering $X$ by analytical continuation to
get function $\Psi_0(P\in X)$.
The following statement takes place:
\begin{theorem}
\la{PsiL0}
Let function  $\Psi(P\in X)$ be defined by the formulas (\ref{Phi}),
(\ref{phi1}), (\ref{psi1}),  (\ref{Psi}), and function $\Psi_0(P\in X)$
be defined by formulas  (\ref{Phi0}),  (\ref{phipsi0}), (\ref{Psi00}).
Let the components of vectors $\pb,\qb\in \C^{2g_0-1}$ and
$\rb,\sb\in \C^{g_0}$ be related as follows:
\be
p_1=-\sum_{l=1}^{g_0} r_l\;, \hskip0.5cm 
q_1=-2 s_1\;;\hskip1.0cm\; 
p_j= r_j\;,\hskip0.5cm  
q_j= s_j - s_1 \;,  \hskip0.5cm 
2\leq j\leq g_0
\la{pqrs}\ee
(relations (\ref{pqs}) automatically give other $g_0-1$ components of $\pb$ and $\qb$). Require in addition that $\left[^\rb_\sb\right]$ is not a half-integer
characteristic.

Then functions $\Psi$ and  $\Psi_0$ coincide:
\be
\Psi(P\in X)=\Psi_0(P\in X)\;.
\la{PsiPsi0}\ee

\end{theorem}

{\it Proof.} The functions $\phi_0$ and $\psi_0$ were chosen in such a
way that the analytical continuation of $\Psi_0$ from $\O_\l$ on the whole
universal covering $X$ does not violate reduction restriction (\ref{coset}). 
Taking into account the non-triviality assumption of non-coincidence of
characteristic  $\left[^\rb_\sb\right]$ with any  half-integer
characteristic, and 
coincidence of the normalization conditions of  $\Psi$ and  $\Psi_0$ at
$\g=\infty$, it is enough to show that monodromies of $\Psi$ and  $\Psi_0$
along the first $g_0$ pairs of basic cycles of curve $\L$ coincide.
For our choice of the basic cycles on $\L$ and $\L_0$ we get the
following relations between the basic cycles
of $\L_0$ and $\L$:
\ben
\Pi(a_1)=-(a_1^0+\dots+a_{g_0}^0)\;,\hskip1.0cm
\Pi(b_1)=-2b_1^0\;,\hskip1.0cm
\Pi(a_j)=a^0_j\;,\hskip1.0cm
\Pi(b_j)=b^0_j-b_1^0\;,
\een
which imply the following
expressions for  monodromies of $\Psi_0$ around $(a_j,b_j)$:
\be
M^0_{a_1}=\exp\{-2\pi i\sum_{j=1}^{g_0}r_j\sigma_3\}  \;,\hskip1.0cm
M^0_{b_1}=\exp\{4\pi i s_1\sigma_3\}
\la{Ma1j}\ee
\be
M^0_{a_j}=  \exp\{2\pi r_j\sigma_3\}\hskip1.0cm 
M^0_{b_j}=  \exp\{2\pi (s_1-s_j)\sigma_3\}\;,\hskip0.5cm 2\leq j\leq g_0
\la{Mb1j}\ee
coinciding with monodromies of  $\Psi$ (\ref{MaMb}) provided
conditions (\ref{pqrs}) are fulfilled (reduction (\ref{coset})
provides coicidence of remaining  monodromies around 
$(a_j,b_j), \; g_0+1\leq j\leq 2g_0$). [Notice that functions $\phi_0$ and 
$\psi_0$ transform in a different way (their monodromies differ by sign) along the basic cycles of $\L_0$,
but their pullbacks on $\Lh$ tranform in the same way along the basic cycles of $\L$.]
\vskip0.5cm
Now we are in position to formulate the following statement:
\begin{theorem}
\la{main}
Let  $\rb,\sb\in \C^{g_0}$ be arbitrary  constant vectors satisfying 
reality conditions (\ref{abj}), (\ref{bbj}). 
Then the following function:
\be
\E(\xi,\xb)=
\frac{\Th\left[^\rb_\sb\right]\left(V\big|_\xi^{\infty^1} \Big|\B_0\right)}
{\Th\left[^\rb_\sb\right]\left(V\big|_\xi^{\infty^2}\Big|\B_0\right)}\;,
\la{E0}\ee
 solves the  Ernst equation (\ref{EE}).
Function $\Psi$ defined equivalently by (\ref{Psi}) or (\ref{Psi00})
satisfies the linear system (\ref{lsg}) with matrix $G(\xi,\xb)$  given
by (\ref{g}), and 
\be
\Psi(\g=\infty)= G
\la{Psig}\ee
\end{theorem}

{\it Proof.} The non-trivial part is to check (\ref{E0}).
One can represent matrix $\Psi(\l=\infty^1)$,  given by (\ref{Psi00}),
in the form (\ref{g}) with
\ben
\E=-i\f{\phi_0(\infty^1)}{\phi_0(\infty^2)}
\een
which leads to  (\ref{E0}) after substitution of (\ref{phipsi0}).
\vskip0.5cm

Now it arises the non-trivial problem of integration of equations (\ref{coeff})
for the metric coefficients $F$ and $k$.
The following statement shows how to find 
the metric coefficient $F(\xi,\xb)$ corresponding to the Ernst potential
(\ref{E0}):
\begin{theorem}
\la{Atheo}
The metric coefficient $F(\xi,\xb)$ (\ref{coeff}) corresponding to the 
Ernst potential (\ref{E0}) is given by 
\be
F=\f{2}{\Re\E}\Im\Big\{\sum_{j=1}^{g_0}({\cal A}_0^{-1})_{g_0 j}\f{\p}{\p z_j}
\log\Th\left[^\rb_\sb\right]\left(V\big|_\xi^{\infty^2} \Big|\B_0\right)
\Big\}
\la{AAA}\ee
up to an arbitrary additive constant,
where ${\cal A}_0$ is the matrix of $a$-periods of holomorphic 1-forms (\ref{A0}); $\f{\p\Th}{\p z_j}$ denotes derivative of theta-function with 
respect to its $j$th argument.
\end{theorem}
{\it Proof.}
Consider the following simple identity
\be
(\Psi^{-1}\Psi_{\l^{-1}})_{\xi}\equiv\Psi^{-1}\left(\Psi_{\xi}\Psi^{-1}\right)_{1/\l}\Psi
\la{ident}\ee
Evaluation  of (12) matrix element of 
the right hand side of (\ref{ident}) at $\g=\infty$ using the
linear system (\ref{lsg}), equation for $F$ (\ref{coeff}) and 
normalization condition (\ref{norm}) gives
nothing but $\f{1}{2} F_{\xi}$. 
Therefore, we can integrate equations for $F$ to get
\be
F=2\Big(\Psi^{-1}\Psi_{\l^{-1}}\Big)_{12}(\g=\infty)
\la{Fgen}\ee
Substitution of
(\ref{phipsi0}), (\ref{phipsi01}) and (\ref{Psi00}) into (\ref{Fgen}) leads to the following expression:
\ben
F=-\f{2}{\Re\E}\Im\Big\{\f{dW(P)}{d\f{1}{\l}}(\l=\infty^2)\Big\}
\een
where the 1-form $dW_0(P\in\L_0)$ is given by
\ben
dW_0(P)=\sum_{j=1}^{g_0} d V_j (P)
\f{\p}{\p z_j}
\log\Th\left[^\rb_\sb\right]\left(V\big|_\xi^{P} \Big|\B_0\right)
\een 
Using formula (\ref{dVj}) for
basic differentials $dV_j$ , we come to (\ref{AAA}).
\vskip0.5cm
The next theorem shows how to integrate
equation (\ref{coeff}) for  remaining metric coefficient $k$: 
\begin{theorem}
The metric coefficient $e^{2k(\xi,\xb)}$, solving equation (\ref{coeff})
with  Ernst potential defined by  (\ref{E0}), 
is given by the following expression:
\be
e^{2k}= 
\f{\Th\left[^\pb_\qb\right]\left(0 |\B\right)}
{\sqrt{\det {\cal A}_0}}\prod_{j=1}^{2g_0}|\l_j-\xi|^{-1/4}
\la{confac}\ee
up to multiplication with an arbitrary 
constant, where $\B$ is the matrix of $b$-periods of curve $\L$ (\ref{L}); 
matrix ${\cal A}_0$ of $a$-periods of holomorphic differentials on the curve $\L_0$  is defined by (\ref{A0}); the constant vectors $\pb,\qb\in \C^{2g_0-1}$ are expressed via
(\ref{pqrs}) in terms of vectors $\rb,\sb\in\C^{g_0}$.
\la{ktheo}
\end{theorem}
{\it Proof.} One should exploit the  coincidence
of functions $\Psi_0$ and $\Psi$,
and substitute the formula for $\tau$-function 
of the Schlesinger system (\ref{tau}) 
into the relation  (\ref{taucon}) between the $\tau$-function and the 
coefficient $e^{2k}$ (in the present case 
$\tr A_j^2=1/8$). 
In addition we have to use the following relation between
determinants of matrices of $a$-periods of holomorphic 1-forms
 of curves $\L$ and $\L_0$:
\be
\det {\cal A}=const\;\rho^{g_0^2}\det {\cal A}_0
\la{AA0}\ee
where the constant is independent of $(\xi,\xb)$.
Proof of (\ref{AA0}) may be obtained by elementary manipulations
using decomposition of holomorphic 1-forms on $\L$ into  combination
of holomorphic 1-forms on $\L_0$ and holomorphic 1-forms on 
$(\xi,\xb)$-independent hyperelliptic curve of  genus $g_0-1$
defined by the equation
\be
\delta^2=\prod_{j=1}^{2g_0}(\l-\l_j)\;.
\la{L1}\ee

To complete the proof of (\ref{confac}) it remains to make use of the folowing relations:
\ben
\g^{-1}\g_\l=\{(\l-\xi)(\l-\xb)\}^{-1/2}\;;
\hskip0.7cm
\prod_{j<k} (\g_j-\g_k)=const\;\rho^{-4g_0^2}\prod_{j=1}^{2g_0}
\{(\l_j-\xi)(\l_j-\xb)\}^{1/2}\;.
\een
\vskip0.5cm

\begin{remark}\rm
Using the standard expression for Vandermonde determinant, we  
get the following formula for $\det {\cal A}_0$:
\be
\det {\cal A}_0 =  \oint_{a_1}\dots\oint_{a_{g_0}}\f{\prod_{k<j} (\mu_j-\mu_k)}
{\prod_{j=1}^{g_0}\nu(\mu_j)}d\mu_1\dots d\mu_{g_0}
\la{detA0}\ee
\end{remark}
\begin{remark}\rm
The theta-function $\Th\left[^\pb_\qb\right]\left(0 |\B\right)$ from 
(\ref{confac}) may be
decomposed into combination of theta-functions associated to 
double matrices of $b$-periods $2\B_0$ and $2\B_1$ of
curves  (\ref{L0})
and (\ref{L1}) (see \c{Fay}). 
\end{remark}
\begin{remark}\rm
Half-integer $\rb$ and $\sb$.
\end{remark}
\begin{remark}\rm
Degeneration $\l_{2j+1},\;\l_{2j+1}\to\ka_j\in\R$ of the spectral curve $\L_0$ 
gives after appropriate choice of $[\rb,\sb]$ the family of multi-Kerr-NUT
solutions. In particular, to get Kerr-NUT soultion itself one has to take
$g_0=2$ and
\ben
r_j=\f{n_j}{2}\;;\hskip0.5cm s_j=\f{n_j}{4}+i\alpha_j,\hskip0.5cm 
\alpha_j\in \R\;, \hskip0.5cm n_j=\pm 1\;, \hskip0.5cm j=1,2
\een
Then (\ref{E0}) turns into \c{Koro88}:
\ben
\E=\f{1-\Gamma}{1+\Gamma}\;, \hskip0.7cm
\Gamma^{-1}=\f{i(a_1-a_2)}{1-a_1 a_2 +i(a_1+a_2)}X+
\f{1+ a_1 a_2}{1-a_1 a_2 +i (a_1+a_2)}Y
\een
where $a_j= n_j e^{-2\pi i \alpha_j}$, $j=1,2$, and 
\ben
X=\f{1}{\ka_1-\ka_2}
\Big\{\sqrt{(\ka_1-\xi)(\ka_1-\xb)}+\sqrt{(\ka_2-\xi)(\ka_2-\xb)}\Big\}
\een
\ben
Y=\f{1}{\ka_1-\ka_2}
\Big\{\sqrt{(\ka_1-\xi)(\ka_1-\xb)}-\sqrt{(\ka_2-\xi)(\ka_2-\xb)}\Big\}
\een
are prolate ellipsoidal coordinates.
\end{remark}
\begin{remark}\rm
Elliptic case: $r=0$ - ``toron'' solution; $r\neq 0$-non-asymptotically flat.

\end{remark}
\section{Known form of algebro-geometric solutions of Ernst equation} 
\setcounter{equation}{0}

The original algebro-geometric solution of Ernst equation 
found in \c{Koro88} looks as follows in 
notations introduced in \c{Koro93,Koro96}: 
\be
\E=\f{\Th\left(V\big|_\xi^{\infty^1} + B_W\Big|\B_0\right)}
{\Th\left(V\big|_\xi^{\infty^2} + B_W\Big|\B_0\right)} 
\exp\left\{W\big|_{\infty^2}^{\infty^1} \right\}
\la{PRA}\ee
where $dW(P)$ is an  arbitrary 1-form on $\L_0$ 
satisfying the following conditions:
\begin{enumerate}\rm
\item
$dW$ is an arbitrary finite, infinite or continuous linear 
combination, with $(\xi,\xb)$-independent coefficients, 
of normalized (all $a$-periods vanish)  
Abelian differentials on $\L_0$ of the 2nd and 3rd kind 
with  $(\xi,\xb)$-independent poles and singular parts.
\item
Reality condition:
\be
dW(\bar{P})=\overline{dW(P)}\;,\hskip1.0cm 2\pi i B_W\in \R^{g_0}\;.
\la{rcdW}\ee

\end{enumerate}
Vector $2\pi iB_W$ is the vector of $b$-periods of
differential $dW$:
\be
2\pi i (B_W)_j=\oint_{b_j}dW(P)
\la{BW}\ee

\subsection{Role of 1-form $dW$}
\subsubsection{Static solutions ($g_0=0$)}

To clarify the role of differential $dW$ we shall consider first the 
simplest case $g=0$, when (\ref{PRA}) becomes real:
\be
\log\E=W\big|_{\infty^2}^{\infty^1} 
\la{static}\ee 
which is real as a corollary of (\ref{rcdW}),
and the Ernst equation linearises to the Euler-Darboux equation:
\be
(\p_\rho^2+\f{1}{\rho}\p_\rho+\p_z^2)\log\E=0
\la{ED}\ee
The natural question which was asked in \c{Koro93} is whether it
is possible to represent general solution of (\ref{ED})
invariant with respect to the change of sign of $\rho$ in the form
(\ref{static}). The positive  answer was given in \c{Koro96}.

Namely, if we fix some domain $D$ in $(x,\rho)$-plane, 
symmetric with respect to involution $\rho\to -\rho$,
(which may contain the infinite point with vanishing boundary condition
$\log \E\underset{x,\rho\to\infty}\to 0$), the general solution
of (\ref{ED}) in this domain may be represented by  the contour
 integral over boundary $\p D$ (which is nothing but axisymmetric version
of  familiar representation of arbitrary static 
electric field inside of a shell
as an induced field of certain charge distribution on the shell):
\be
\log \E =
\int_{\p D} \f{f(\ka)d\ka}{\{(\ka-\x)(\ka-\xb)\}^{1/2}}\;,
\la{SB}
\ee
with arbitrary function $f(\ka)$ defined on $\p D$ and satisfying 
reality condition
\ben
f(\bar{\ka})=\overline{f(\ka)}\;.
\een

%where 
%\be
%h(\mu)=-f'(\mu)\;.
%\la{hf}\ee

Solution (\ref{SB}) may be represented in the form (\ref{static}), 
where $dW$ is some locally holomorphic 1-form on the rational curve $\L_0$ 
(\ref{L0}) with $g_0=0$.  For that one should take 
\be
dW(\l)=\f{1}{2}\oint_{\p D} f(\ka) (dW_{\ka}(\l)-dW_{\ka^*}(\l)) d\ka\;,
\la{dW1}\ee
where $dW_\ka(\l)$ is differential of the 2nd kind
on $\L_0$ with unique simple pole
at $\l=\ka$ and the following local expansion at $\l=\ka$:
\be
dW_\ka(\l)\underset{\l\to\ka}=\left(\f{1}{(\l-\ka)^2}+ O(1)\right)d\l\;.
\la{locdWm}\ee
In the present $g_0=0$ case we can write $dW_\ka (\l)$ explicitly
as follows:
\be
dW_\ka (\l)=\f{\g'(\ka)\g'(\l)d\l}{(\g(\l)-\g(\ka))^2}\;,
\la{dWm1}\ee
and, therefore,
\be
\int_{\infty^1}^{\infty^2} dW_\ka =\p_\ka (\log\g)\big|_{\g=\infty}^{\g=0}
=\f{1}{\{(\ka-\x)(\ka-\xb)\}^{1/2}}\;,
\la{W1i1i2}\ee
which shows coincidence of  (\ref{static}) with  (\ref{SB}).

The  partial integration in (\ref{dW1}) leads to the
following alternative representation of $dW$:
\be
dW(\l)=\oint_{\p D} h(\ka) dW_{\ka\,\ka^*}(\l) d\ka\;,
\la{dWl2}\ee
where  
\be
h(\ka)=-\f{f'(\ka)}{2}
\la{hf}\ee
and
$dW_{\ka\,\ka^*}(\l)$ is the standard differential of the 3rd kind
given by
\be
dW_{\ka\,\ka^*}(\l)=\left(\f{\g'(\l)}{\g(\l)-\g(\ka)}-
\f{\g'(\l)}{\g(\l)-\g^{-1}(\ka)}\right)d\l\;.
\la{dWm2}\ee
Integral representation (\ref{dWl2}) of $dW$ is the counterpart of the  
partially integrated version of (\ref{SB}):
\ben
\log \E =
2\int_{\p D} h(\ka)\log \g(\ka,\xi,\xb)d\ka\;.
\een

Performing in (\ref{dW1}) the partial integration in a different way we
can represent $dW$ as a contour integral over $\ka$ of meromorphic
1-forms having poles of arbitrary fixed order at $\l=\ka$ and
$\l=\ka^*$. We see that from the local point of view the set of 
1-forms $dW$ allowed in
(\ref{static}) is over-complete, and in order to get general local static 
solution it is sufficient to restrict ourselves by forms $dW$  
represented by (\ref{dWl2}).

\subsubsection{Forms $dW$ for higher genus}

Denote by $dW_{QR}(P)$ the differential of the 3rd kind on $\L_0$ 
with poles of the first order at $Q$ and $R$ and residues $+1$ and
$-1$ respectively, satisfying normalization conditions
\be
\oint_{a_j} dW_{QR}(P)=0\;,\hskip0.5cm \forall j\;.
\la{normdWQR}\ee
The vector of $b$-periods of $dW_{QR}$:
\ben
2\pi i (B_{QR})_j\equiv \oint_{b_j} dW_{QR}
\een 
may be expressed in the following way in terms of basic holomorphic 1-forms
$dV_j$:
\be
(B_{QR})_j=\int_R^Q dV_j\;.
\la{BQR}\ee

For $g_0\geq 1$ it takes place the situation similar to $g_0=0$: 
locally the same 1-form $dW(P\in\L_0)$ can be represented in many different ways as contour integral of elementary differentials.
The representation which it will be convenient to use in this paper
is a $g_0\geq 1$ counterpart of (\ref{dWl2}):
\be
dW=\oint_{\p D}h(\ka)dW_{\ka\,\ka^*} \;d\ka\;.
\la{dWg0}\ee
For vector of $b$-periods we have
\be
(B_W)_j=2\oint_{\p D}h(\ka)V_j\big|_\xi^\ka  d\ka\;.
\la{BWg0}\ee
Now one can easily see the link to papers \c{KleRic97},
where it was noticed that certain combination of
 components of the function $\Psi$ from (\ref{ls}) corresponding to solution (\ref{PRA}), 
(\ref{dWg0}), solves scalar Riemann-Hilbert problem on curve $\L_0$
with contour $\p D$. Partial integration in (\ref{dWg0}), (\ref{BWg0}) gives
the following relation between our function $h(\ka)$ and conjugation 
function of RH problem $G(\ka)$ from \c{KleRic97}:
\be
\p_\ka\log G(\ka)= - 4\pi i h(\ka)\;.
\la{Gh}\ee
The proof of (\ref{Gh}) can be obtained expressing corresponding objects
in terms of the prime-form on $\L_0$.

Now we are going to discuss the relationship between solutions (\ref{E0}),
which we derived exploiting the link between Ernst and Schlesinger equations,
and previuosly known class of solutions of Ernst equation (\ref{PRA}). 
Apparently, solutions (\ref{E0}) contain less parameters; however, 
it turns out that in some sense both classes of solutions coincide.

\subsection{Solutions (\ref{E0}) from solutions (\ref{PRA})}

To describe this link we shall briefly discuss how 
abelian differentials of the 1st kind on $\L_0$ arise as limits
of abelian differentials of the 3rd kind.

\subsubsection{Holomorphic 1-forms as limits of meromorhic 1-forms}

Cut $\L$ along basic cycles started at the same point and 
consider $dW_{QR}(P)$ in  the fundamental polygon $\Lh_0$ whose boundary
consists of the cycles ${a_j^0}^+,\,{b_j^0}^+,\,{a_j^0}^-,\,{b_j^0}^+$. 

Obviously, in the limit $Q\to R$ differential $dW_{QR}$ should turn into
certain holomorphic 1-form.
The most naive way to perform this limit is to take $Q\to R$ 
inside of $\Lh_0$; then 
 $dW_{QR}\to 0$ and $B_{QR}\to 0$. However, this limit may also be taken 
in two different non-trivial ways:
\begin{enumerate}
\item
Let $Q$ tends to some boundary point of $\Lh_0$, say, to a point 
belonging to the contour ${b_j^0}^+$; and let $R$ tend simultaneously to 
corresponding point of the contour ${b_j^0}^-$. 
Then the limiting procedure 
does not influence the normalization conditions (\ref{normdWQR}), and, 
since in the limit  $dW_{QR}$ turns into holomorphic 1-form, this
1-form must vanish:
\be
dW_{QR}\underset{Q\to R\in b_j^0 }\to 0\;.
\la{limdW1}\ee
 However, the vector of $b$-periods of  $dW_{QR}$ does not
vanish even in the limit: from (\ref{BQR}) we see that 
\be
(B_{QR})_k\underset{Q\to R\in b_j^0}\to -\delta_{jk}\;.
\la{limB1}\ee

\item
Alternatively, we can treat the limit assuming that
 $Q$ tends to some point of $\Lh_0$
on the contour ${a_j^0}^+$, and $R$ tends to the same point 
of $\L_0$ 
belonging to the contour  ${a_j^0}^-$. Since the poles of $dW_{QR}$ meet
exactly on the cycle $a_j^0$, we see that
the  integral $\oint_{a_j^0}dW_{QR}$ does not vanish in the limit any more. 
Since  $\oint_{a_k^0}dW_{QR}$ still vanishes for all $k\neq j$, we conclude
that in the limit  $dW_{QR}$ becomes proportional to $dV_j$. To find 
the coefficient of proportionality, we calculate vector of $b$-periods of
 $ dW_{QR}$ in the limit according to  (\ref{BQR}) and find that 
\be
B_{QR}\underset{Q\to R\in a_j^0}\to {\B_0}_j\;,
\la{limB2}\ee
where ${\B_0}_j$ stands for $j$th column of the matrix of 
$b$-periods. Therefore, for $dW_{QR}$ itself we have
\be
dW_{QR}\underset{Q\to R\in a_j^0}\to 2\pi i\, dV_j\;.
\la{limdW2}\ee 
\end{enumerate}

Now choose differential $dW$ in solution (\ref{PRA}) in the following form: 
\be
dW = \sum_{j=1}^{g_0} (r_j dW_{Q_j R_j} - s_j dW_{\Qt_j \Rt_j})
\la{dWSchl}\ee
with some constant vectors $\rb,\sb\in\C^{g_0}$,
and take the limits  $Q_j\to R_j\in a_j^0$,  $\Qt_j\to \Rt_j\in b_j^0$
as explained in items 1 and 2 above. 
In this limit, according to (\ref{limdW1}), (\ref{limB1}), (\ref{limB2})
 (\ref{limdW2}), we get
\be
B_W\to \B_0\rb +\sb\;,
\la{Bschl}\ee
\be
W\big|_{\infty^2}^{\infty^1} \to \sum_{j=1}^{g_0}r_j V_j\big|_{\infty^2}^{\infty^1}\;.
\la{Wschl}\ee
According to reality conditions (\ref{rcdW}) we should impose
the following restriction on constants $\rb$ and $\sb$:
\ben
\B_0\rb +\sb\in i\R^{g_0}\;,
\een
coinciding with (\ref{rcalt}). Therefore,  solution (\ref{PRA}) turns
in this limit into
\ben
\E=\f{\Th\left( V\big|_\xi^{\infty^1} + \B_0\rb+\sb\Big|\B_0\right)}
{\Th\left( V\big|_\xi^{\infty^2} + \B_0\rb+\sb\Big|\B_0\right)}
\exp\Big\{ \sum_{j=1}^{g_0} r_j V_j\big|_{\infty^2}^{\infty^1} \Big\}\;,
\een
which coincides with (\ref{E0}) if we take into account that the theta-function
with characteristics in nothing but the ordinary theta-function with
shifted argument multiplied with certain exponential factor (\ref{charac}).

Next we shall consider more non-trivial procedure of coming from
(\ref{E0}) to (\ref{PRA}), (\ref{dWg0}).

\section{General algebro-geometric solutions of Ernst equation as limits
of Schlesinger-related ones}
\setcounter{equation}{0}

Here we shall describe the inverse procedure: how to get the  class
of solutions (\ref{PRA}), (\ref{dWg0}) starting from solutions (\ref{E0}).

\subsection{Partial degeneration of spectral curve}

Let us consider solution (\ref{E0}) with curve $\L_0$ substituted by 
the curve $\L_1$ of genus $g_1=g_0+n$ defined by the equation
\be
\nu^2=(\l-\xi)(\l-\xb)\prod_{j=1}^{2g_0+2n} (\l-\l_j)\;,
\la{L10}\ee
and choose  the vectors $\rb,\,\sb\in \C^{g_0+n}$ in the following way:
\be
\sb=0 \;;\hskip0.5cm r_j=0\;,\hskip0.5cm 1\leq j\leq g_0\;;
\hskip0.5cm r_{j+g_0}= h_j\in \R\;,\hskip0.5cm 1\leq j\leq n\;.
\la{rsdeg}\ee 
Without loss of generality we shall assume $h_j\in [0,1]$.
Denote by $\B_1$ the matrix of $b$-periods of $\L_1$ and
by $dV_1,\dots, dV_{g_0+n}$ the basis of normalized holomorphic 1-forms
on $\L_1$.
Now consider the  solution (\ref{E0}) constructed from these data: 
\be
\E=\f{\Th\left[^\rb_0\right]\left(V\big|_\xi^{\infty^1}\Big|\B_1\right)}
{\Th\left[^\rb_0\right]\left( V\big|_\xi^{\infty^2} \Big|\B_1\right)}\;,
\la{E1}\ee
and take the limit 
\be
\l_{2g_0+2j+1}\;,\l_{2g_0+2j+2}\to \ka_j\in\R\;.
\la{deg}\ee
Then curve $\L_1$ turns into $\L_0$ with 
double points at $\ka_j\;,j=1,\dots,n$.
The basis of holomorphic 1-forms of $\L_1$ turns into 
\ben
dV_1,\dots,dV_{g_0},\f{1}{2\pi i}
dW_{\ka_1\ka_1^*},\dots,\f{1}{2\pi i}dW_{\ka_n\ka_n^*}\;,
\een
where $dV_1,\dots,dV_{g_0}$ is the basis of normalized holomorphic 1-forms
on $\L_0$, and $dW_{\ka_j\ka_j^*}$ are normalized 1-forms of the 
3rd kind  on $\L_0$ with  simple poles at $\ka_j$ and $\ka^*_j$
and residues $+1$ and $-1$ respectively.
Therefore, the  matrix of $b$-periods of $\L_1$ 
in the limit (\ref{deg}) may be described in terms of the objects associated to
the curve $\L_0$  as follows:
\be
(\B_1)_{jk}= (\B_0)_{jk} + o(1)\;,\hskip0.5cm 1\leq j,k\leq g_0\;;
\la{B1B0}\ee
\be
(\B_1)_{j\; k+g_0} = 2V_j\big|_\xi^{\ka_k}  + o(1)\;,\hskip0.5cm  
1\leq j\leq g_0\;,\hskip0.5cm 1\leq k\leq n\;;
\la{B1dW}\ee
\be
(\B_1)_{j+g_0\;j+g_0} = -\f{1}{\pi i}
\log |\l_{2g_0+2j+1}-\l_{2g_0+2j+2}|+ O(1)\;,\hskip0.5cm 1\leq j\leq n\;;
\la{Bdiag}\ee
\ben
(\B_1)_{k+g_0\;j+g_0}=O(1)\;,\hskip1.0cm 1\leq j\neq k\leq n\;.
\een
Substituting  these relations in the definition of theta-function
 we can  express the value of Ernst potential (\ref{E1}) 
in the limit (\ref{deg}) in terms of the objects associated to
the curve $\L_0$: 
\be
\E=\f{\Th\left( V\big|_\xi^{\infty^1} +2\sum_{j=1}^n h_j V\big|_\xi^{\ka_j}
\Big|\B_0\right)}{\Th\left( V\big|_\xi^{\infty^2}+2\sum_{j=1}^n h_j V\big|_{\xi}^{\ka_j}\Big|\B_0\right)}
\exp\Big\{\sum_{j=1}^n h_j W_{\ka_j\ka_j^*}\big|_{\infty^2}^{\infty^1}\Big\}
\la{degE1}\ee

The formula (\ref{AAA})  for the metric coefficient $F$ transforms into
\be
F= \f{2}{\Re\E}\Im\Big\{\sum_{j=1}^{n}({\cal A}_0^{-1})_{g_0 j}\f{\p}{\p z_j}
\log\Th\Big(V\big|_\xi^{\infty^2}+2\sum_{j=1}^{n}h_j V\big|_\xi^{\ka_j} \Big|\B_0\Big)
+\sum_{j=1}^{n}h_j\f{dW_{\ka_j \ka^*_j}}{d\l^{-1}}(\infty^2)
\Big\}
\la{Fdis}
\ee
For coefficient $e^{2k}$ we get from (\ref{confac}):
\ben
e^{2k}=\{\det {\cal A}_0\}^{-\f{1}{2}}\prod_{j=1}^{2g_0}|\l_j-\xi|^{-\f{1}{4}}
\Th\Big(\sum_{j=1}^{n}h_j(U\big|_1^{\g (\ka_j)} - 
\eb_1)\Big|\B\Big)
\een
\be
\times\exp\Big\{2\pi i\sum_{j=1}^{n}h_j^2\beta_j+
\pi i \B_{11}\Big(\sum_{j=1}^n h_j\Big)^2+
\sum_{j\neq k,\;j,k=1}^{n}h_j h_k W_{\ka_k\ka_k^*}\big|_{\ka_j^*}^{\ka_j} 
\Big\};,
\la{kdis}\ee
where as before $\B$ is $2g_0-1\times 2g_0-1$ matrix of $b$-periods of  
curve $\L$;
$\beta_j$ is the second term of asymptotical expansion of
${(\B_1)}_{g+j\;g+j}$ as $\l_{g+2j+1}\to\l_{g+2j+2}$ ($j=1,\dots,n$):
\ben
(\B_1)_{g+j\;g+j}=\f{1}{\pi i}\log|\l_{g+2j+1}-\l_{g+2j+2}|+\beta_j+o(1)\;.
\een

\begin{remark}\rm
The assumption $\ka_j\in\R\;,h_j\in\R$ 
was only made to shorten the presentation;  one can also
allow in (\ref{degE1}) the presence of
conjugated pairs $\ka_j=\bar{\ka}_l$, $h_j=\bar{h}_l$ which come out
if we glue together  two ``vertical'' branch cuts.
\end{remark}

\subsection{Continuous limit: condensation of double points}

Now we can take a continuous limit in (\ref{degE1}) distributing
points $\ka_j$ over an arbitrary contour $\p D$ with an arbitrary
(say, continuous) measure 
$h(\ka)$ satisfying reality condition
\ben
h(\bar{\ka})=\overline{h(\ka)}\;.
\een
Then (\ref{degE1}) turns into
\be
\E=\f{\Th\left( V\big|_\xi^{\infty^1} +2\oint_{\p D} h(\ka) V\big|_\xi^{\ka}d\ka
\Big|\B_0\right)}{\Th\left( V\big|_\xi^{\infty^2} +2\oint_{\p D} h(\ka) V\big|_\xi^{\ka}d\ka\Big|\B_0\right)}
\exp\left\{ \oint_{\p D}h(\ka) W_{\ka\ka^*}\big|_{\infty^2}^{\infty^1}d\ka\right\}
\la{degEcon}\ee
coinciding with (\ref{PRA}), (\ref{dWg0}), (\ref{BWg0}).

Taking continuous limit in the formulas (\ref{Fdis}), (\ref{kdis}),
we come to the following expressions for the metric coefficients $e^{2k}$ 
and $F$ corresponding to the Ernst potential (\ref{degEcon}).

\begin{theorem}
Coefficient $F$ of the metric (\ref{metric}), corresponding to solution
(\ref{degEcon}) of the Ernst equation, is given by
\ben
F=
\een
\be
\f{2}{\Re\E}\Im\Big\{\sum_{j=1}^{g_0}({\cal A}_0^{-1})_{g_0 j}\f{\p}{\p z_j}
\log\Th\left( V\big|_\xi^{\infty^2}+2\oint_{\p D} h(\ka) V\big|_\xi^{\ka}d\ka\Big|\B_0\right)+\oint_{\p D} h(\ka)d\ka\f{dW_{\ka \ka^*}}{d\l^{-1}}(\infty^2)
\Big\}\;,
\la{Fcont}
\ee
where all the objects are associated to curve $\L_0$.
Metric coefficient $e^{2k}$ is, up to an arbitrary
constant factor,  given by the following expression:
\ben
e^{2k}=
\{\det {\cal A}_0\}^{-\f{1}{2}}\prod_{j=1}^{2g_0}|\l_j-\xi|^{-\f{1}{4}}
\Th\left(\oint_{\p D} h(\ka) (U\big|_1^{\g (\ka)} - 
\eb_1)d\ka\Big|\B\right)
\een
\be
\times\exp\left\{\pi i \B_{11}\Big(\oint_{\p D} h(\ka)d\ka\Big)^2+
\oint_{\p D}\oint_{\p D}\left(W_{\kat\kat^*}\big|_{\ka^*}^\ka -
2\log|\ka-\kat|\right)d\ka d\kat\right\}\;,
\la{kcont}
\ee 
where $\B$ is the matrix of $b$-periods and $dU$ is the normalized basis
of holomorphic 1-forms on curve $\L$. 
\end{theorem}
{\it Proof.} Formulas (\ref{Fcont}) and (\ref{kcont}) are direct 
continuous analogs of  (\ref{Fdis}) and (\ref{kdis}), respectively.
Term $2\log|\ka-\kat|$ is subtracted, using the freedom to renormalize
 $e^{2k}$ with an arbitrary $(\xi,\xb)$-independent factor, to achieve 
convergence of the double integral at $\ka=\kat$.

% \section{Conclusions}

\newpage

\end{document}